\documentclass[%
 aip,
 amsmath,amssymb,
 reprint,%
]{revtex4-1}

\usepackage{graphicx}
\usepackage{dcolumn}
\usepackage{bm}

\usepackage[utf8]{inputenc}
\usepackage[T1]{fontenc}
\usepackage{mathptmx}
\usepackage{etoolbox}
\usepackage{textcomp}

\usepackage{amsmath} 
\usepackage{siunitx}
\usepackage{float}

\usepackage{todonotes}

\newcommand{\GaAs}{\ensuremath{\mathrm{GaAs}}}
\newcommand{\AlAs}{\ensuremath{{\mathrm{AlAs}}}}
\newcommand{\AlGaAs}{\ensuremath{\mathrm{AlGaAs}}}
\newcommand{\nm}{\nano\metre}
\newcommand{\um}{\micro\metre}

\makeatletter
\def\@email#1#2{%
 \endgroup
 \patchcmd{\titleblock@produce}
  {\frontmatter@RRAPformat}
  {\frontmatter@RRAPformat{\produce@RRAP{*#1\href{mailto:#2}{#2}}}\frontmatter@RRAPformat}
  {}{}
}%
\makeatother
\begin{document}

\preprint{AIP/123-QED}

\title[Temperature-dependent refractive index of AlGaAs for quantum-photonic devices near the bandgap]{Temperature-dependent refractive index of AlGaAs for quantum-photonic devices near the bandgap}

\author{M. Langer}
\affiliation{Institute for Emerging Electronic Technologies, IFW Dresden, Helmholtzstraße 20, 01069 Dresden, Germany}

\author{S. A. Dhurjati}
\affiliation{Institute for Emerging Electronic Technologies, IFW Dresden, Helmholtzstraße 20, 01069 Dresden, Germany}
\author{M. Bauer}
\affiliation{Institute for Emerging Electronic Technologies, IFW Dresden, Helmholtzstraße 20, 01069 Dresden, Germany}
\author{Y. G. Zena}
\affiliation{Institute for Emerging Electronic Technologies, IFW Dresden, Helmholtzstraße 20, 01069 Dresden, Germany}
\author{A. Rahimi}
\affiliation{Institute for Emerging Electronic Technologies, IFW Dresden, Helmholtzstraße 20, 01069 Dresden, Germany}
\author{R. Bassoli}
\affiliation{Quantum Communication Networks research group, Deutsche Telekom Chair of Communication Networks, Technische Universität Dresden, Dresden, Germany}
\author{F.H.P. Fitzek}
\affiliation{Deutsche Telekom Chair of Communication Networks, Technische Universität Dresden, Dresden, Germany}
\author{O. G. Schmidt}
\affiliation{ Research Center for Materials, Architectures and Integration of Nanomembranes (MAIN), Chemnitz University of Technology, Chemnitz, Germany}
\author{C. Hopfmann}
\affiliation{Quantum Communication Networks research group, Deutsche Telekom Chair of Communication Networks, Technische Universität Dresden, Dresden, Germany}

\date{\today}

\begin{abstract}
We present an experimental method to determine the refractive index of \(\mathrm{Al_{x}Ga_{1-x}As}\) (\(x = 0.0\,\text{-}\,0.5\)) from \qty{300}{\kelvin} to \qty{4}{\kelvin} across the \qty{500}{\nm}–\qty{1100}{\nm} wavelength range. The values are extracted from spectroscopically observed microcavity resonances in thin \AlGaAs{} membranes embedded between fully and partially reflective gold mirrors. Refined Varshni and Paessler models are used to describe temperature-dependent bandgap shifts and material composition. By tracking resonance shifts and benchmarking against finite-difference time-domain simulations, we derive the dispersive optical response with high precision. This yields a quantitatively improved analytical expression for the refractive index of \AlGaAs{} matching the experimental results with a coefficient if determination as high as \(R^2=0.993\), enabling accurate modeling near the band edge at cryogenic temperatures. The method is straightforward and broadly applicable to other semiconductor systems, offering a valuable tool for the design of micro photonic devices such as quantum light sources.

\end{abstract}

\maketitle

\section{\label{sec:introduction}Introduction}

Quantum communication is rapidly emerging as a transformative technology for secure information transfer, leveraging the fundamental principles of quantum mechanics \cite{Gisin2007, Pan2025, Schmaltz2025, Singh2025}. Achieving practical quantum communication networks that satisfy key performance requirements—such as high throughput, low latency, and scalability—critically relies on the availability of compact and high-fidelity quantum light sources \cite{Semenenko2024, Schmaltz2025}. Among the various candidate platforms, semiconductor quantum dots have proven particularly promising as on-demand sources of nonclassical light \cite{Keil2017, BassoBasset2019, Tomm2021}. In particular, gallium arsenide (GaAs) quantum dots have emerged as leading contenders for generating entangled photon pairs, offering deterministic emission, high brightness, and straightforward integration with fiber-based systems \cite{Chen2000, Akopian2006, Dousse2010, chen2018highly, hopfmann2021maximally, BassoBasset2021, langer2025bright, Margaria2024}. These sources have already enabled important milestones, including fiber-based entanglement distribution and quantum teleportation \cite{Strobel2024, Strobel2024a}.

A key parameter underlying the optical performance of such devices is the refractive index, which fundamentally governs light–matter interactions in semiconductor materials. Accurate knowledge of the refractive index is indispensable for the design and modeling of photonic components such as photonic crystals \cite{uppu2020scalable}, microcavities \cite{rickert2019optimized, shih2023universal}, and microlenses \cite{langer2025bright}. Because the refractive index depends sensitively on factors including material composition, photon energy (or wavelength), temperature, crystal orientation, doping level, and mechanical stress, its precise determination—both theoretically and experimentally—is essential for advancing material understanding and optimizing device performance in quantum photonics.

Spectroscopic ellipsometry \cite{aspnes2014spectroscopic} is the standard technique to characterize the dispersive refractive index of thin films and bulk materials. This method measures the change in the polarization state of light upon reflection from a sample surface. By analyzing the amplitude ratio and phase difference of the reflected light and modeling the resulting parameters, ellipsometry enables the extraction of both the real \(n\) and imaginary \(k\) parts of the complex refractive index.

For materials such as \AlGaAs{}, the refractive index is well documented at room temperature (RT) \(T \geq \qty{300}{\kelvin}\) \cite{adachi1985gaas,aspnes1986optical}. However, accurate measurements at cryogenic temperatures \(T \ll \qty{300}{\kelvin}\) are not available. This is especially true for energies close to the material bandgap, as there the anomalous dispersion effect makes simple extrapolation from ambient temperatures impractical.  Measurements of the cryogenic refractive index are technically challenging, due to the angle-resolved nature of ellipsometry, which becomes impractically in cryogenic environments. Such systems typically have restricted vertical optical access through a window in the cold shield, making the oblique-angle geometry required for ellipsometry difficult to implement. 

To overcome this limitation, previous studies have interpolated low-temperature refractive indices from RT values using bandgap shift models \cite{rickert2019optimized} or have extrapolated distinct values indirectly by modeling optical spectra of integrated devices \cite{bremer2022numerical,chen2018highly,nie2021experimental}. Both approaches lack precision near the material's band edge, and often fail to adequately account for the full compositional, dispersive, and temperature dependence of the refractive index.

In this work, we propose a simplified and effective alternative to conventional ellipsometry that is compatible with the constraints of cryogenic environments. Our method exploits vertical optical probing and does not require angle-dependent measurements. The approach is based on a reflection mode configuration analogous to a Fabry-Pérot interferometer, where discrete optical modes within a microcavity enhance resonant absorption when the optical path length matches an integer multiple of the cavity mode. By analyzing the spectral positions of these cavity resonances and considering the cavity thickness with high precision, we extract the refractive index at the corresponding wavelengths.
This technique allows refractive index measurements across a comprehensive parameter space, including composition \(x=0\,\text{-}\,0.50\), wavelength (from near the band edge to the infrared, limited only by the available spectroscopy apparatus), and temperature \qty{295}{\kelvin} to \qty{4}{\kelvin}.

\section{\label{sec:methods}Methods}
\subsection{\label{sec:methods.1} Sample Fabrication}

Fabrication of microcavities based on suspended \(\mathrm{Al_{x}Ga_{1-x}As}\) membranes enclosed by gold mirrors follows a two-step process. First, the as-grown membranes are released and transferred onto gold-coated \GaAs{} substrates. In the second step, microcavity structures are defined by lithographic patterning and etching.

The heterostructures are grown by molecular beam epitaxy on 3-inch \GaAs{}(100) wafers and comprise a superlattice structure to smoothen the growth front, followed by a \qty{50}{\nm} \GaAs{} buffer layer and a \qty{30}{\nm} \AlAs{} sacrificial layer for membrane release. The active membrane layer consists of \(\mathrm{Al_{x}Ga_{1-x}As}\) with nominal aluminum compositions \(\mathrm{Al_{nom}} = 0.0,\; 0.1,\; 0.15,\; 0.3,\; 0.5\), and \(x\) denoting the actual aluminium content, which may slightly deviate due to growth variability. The nominal thickness of the membrane is \(\mathrm{d_{nom}} = \qty{1800}{\nm}\), including a thin \GaAs{} layers above and below the \AlGaAs{} layer to protect the surfaces during wet chemical processing. For \(\mathrm{Al_{nom}} \leq 0.3\), a \qty{2}{\nm} cap is sufficient, while a higher aluminum content (\(\mathrm{Al_{nom}} = 0.5\)) requires a thicker \qty{10}{\nm} cap to mitigate degradation arising from reduced etch selectivity. In this case, degradation is not due to complete penetration of the etchant through the cap, but rather to dislocation-mediated etch channels formed within the \GaAs{} layer that facilitate attack on the underlying AlGaAs{} membrane. To benchmark reproducibility and demonstrate compatibility with the fabrication of quantum photonic devices, a reference membrane with \(\mathrm{Al_{nom}} = 0.15\) and \(\mathrm{d_{nom}} = \qty{1460}{\nm}\), previously used in microlens-based quantum photonic platforms~\cite{langer2025bright}, is processed in parallel.

The membranes are transferred to \GaAs{} substrates pre-coated with \qty{100}{\nm} thick gold mirrors, following the procedure described in Ref.~\cite{langer2025bright}. A \qty{25}{\percent} aqueous solution of hydrofluoric acid is used to selectively etch \AlAs{} over \GaAs{}. The etching selectivity of \(\mathrm{Al_{x}Ga_{1-x}As}\)\cite{kumar2007sacrificial} is a critical factor in the design of our heterostructures. The sacrificial layer must exhibit the highest etch rate, ideally by at least one order of magnitude, to ensure efficient removal. For \(\mathrm{Al_{nom}} = 0.5\), the non-linear etch selectivity becomes insufficient, resulting in etch attacks on the membranes through the \GaAs{} protection layer, despite its increased thickness. Although further increasing the thickness of the protection layer could mitigate this problem, it would introduce a larger fraction \GaAs{} within the membrane, which would complicate later analyses. For this reason, the thickness of the protection layer should be kept minimal. This limitation results in a significant decrease in the fabrication yield for \(\mathrm{Al_{nom}} \geq 0.5\), highlighting a key constraint in the current fabrication process.

\begin{figure}[htbp]
\centering
\includegraphics[width=\linewidth]{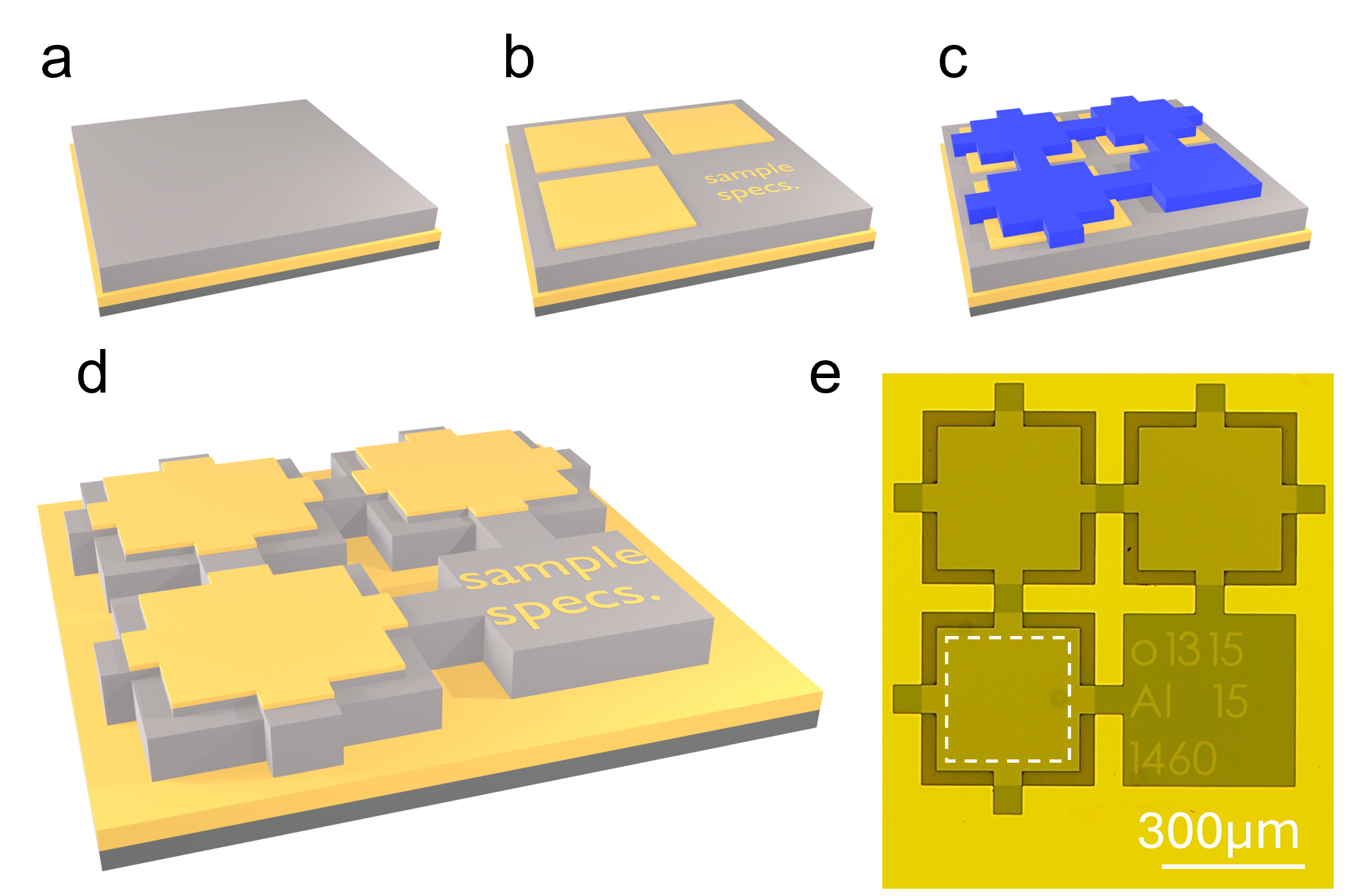}
\caption{\label{fig:fabrication}
Fabrication of planar micro-cavities. a) Membrane on a \qty{100}{\nm} gold mirror. b) Deposition of semi-reflecting gold squares (\qty{350}{\um} lateral, \qty{10}{\nm} thick). For every forth square the top gold layer deposition is omitted, this enables the accurate AFM membrane thickness measurements shown in Tab. \ref{tab:height}. c) Photolithographic patterning with \qty{300}{\um} resist squares (blue). d) Pattern transfer by reactive ion etching, yielding unstrained, free-standing cavities. e) Optical micrograph of the final structure; white dotted square indicating measurement area. }
\end{figure}

Semi-reflective top mirrors are created by electron-beam evaporation coating of \qty{10}{\nm} gold layers patterned into squares with lateral dimensions of \qty{350}{\um}. Photolithographic patterning with \qty{300}{\micro\meter} resist masks defines the membrane regions. This is followed by anisotropic reactive ion etching, which transfers the pattern through the membrane layer to the underlying gold mirror, as illustrated in Fig.~\ref{fig:fabrication}. The gold remains chemically and physically inert and does not undergo etching \cite{langer2025bright}. Accordingly, the effective membrane thicknesses, \(\mathrm{d_{eff}}\), are determined after processing by means of atomic force microscopy (AFM) and are summarized in Table~\ref{tab:height}.

\begin{table}[htbp]
\caption{\label{tab:height} Nominal \(\mathrm{d_{nom}}\) and effective \(\mathrm{d_{eff}}\) membrane-thickness (measured by AFM) with deviation \(\mathrm{dev}\) and nominal aluminum concentration \(\mathrm{Al_{nom}}\). The error of \(\mathrm{dev}\) is the standard deviation of $5$ measurements.}
\begin{ruledtabular}
\begin{tabular}{rrll}
\(\mathrm{Al_{nom}}\)&\(\mathrm{d_{nom}}\) (nm)&\(\mathrm{d_{eff}}\) (nm)&\(\mathrm{dev}\)(\%)\\
\hline
0.00 & 1800 & 1777(7) & -1.3\\
0.10 & 1800 & 1764(5) & -2.0\\
0.15 & 1460 & 1421(5) & -2.6\\
0.15 & 1800 & 1776(31) & -2.4\\
0.30 & 1800 & 1778(20) & -1.2\\
0.50 & 1800 & 1825(5) & +1.4\\
\end{tabular}
\end{ruledtabular}
\end{table}

\subsection{\label{sec:methods.2} Optical characterization}
\begin{figure}[htbp]
\centering
\includegraphics[width=\linewidth]{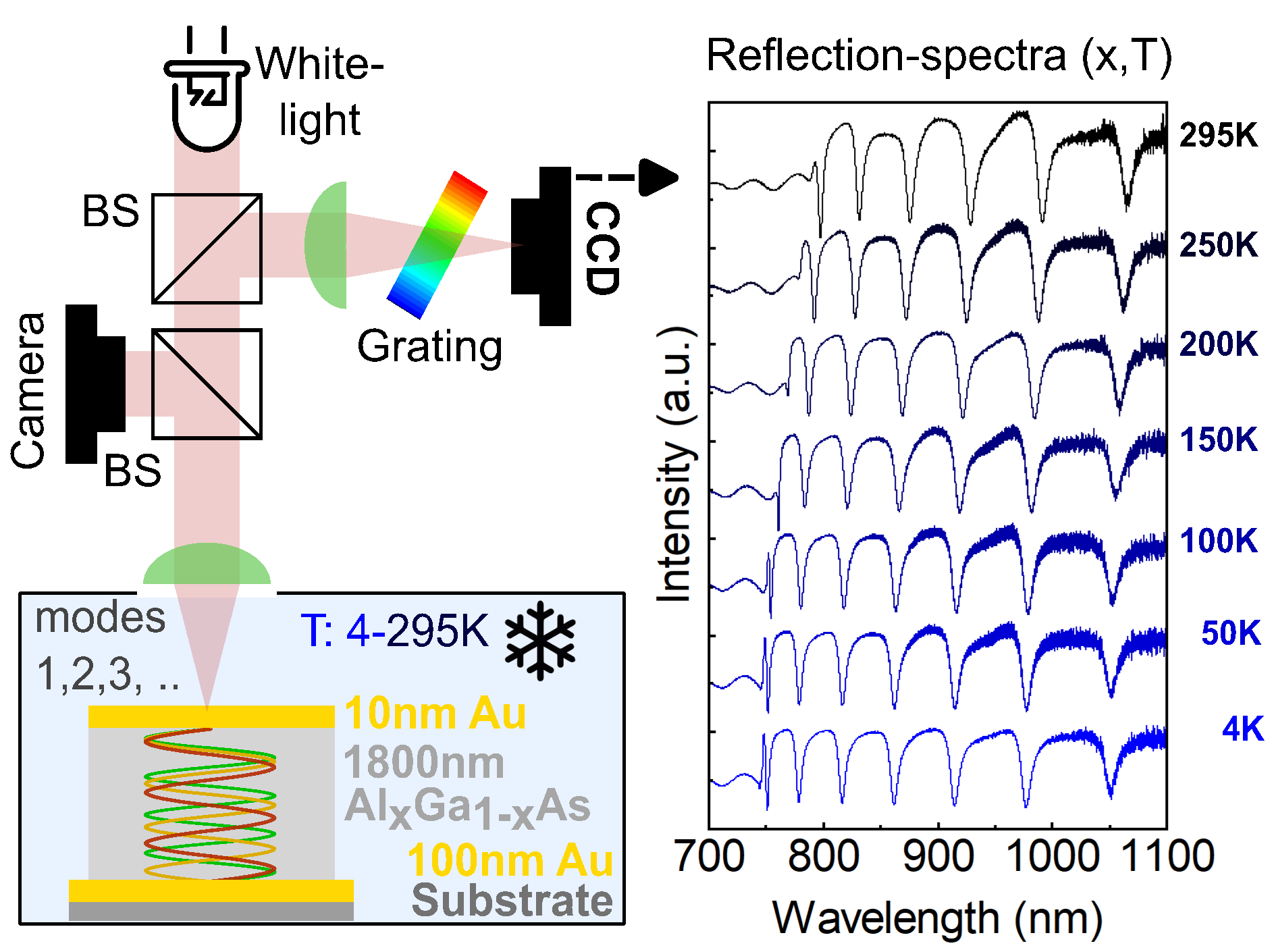}
\caption{\label{fig:setup} White-light reflection spectroscopy at controllable temperature range \qty{295}{}-\qty{4}{\kelvin}. 
Left: Schematic of the experimental setup, with white-light excitation through optical top-window onto membrane sample consisting of an \(\mathrm{Al_{x}Ga_{1-x}As}\) layer enclosed between a fully reflecting (bottom) and a semi-reflecting (top) Au mirror, mounted in a He-flow cryostat with heating element. Live imaging and positioning is performed by a Complementary Metal-Oxide-Semiconductor (CMOS) camera. Reflectance and PL spectra are recorded using a charge couled device (CCD) attached to a \qty{0.5}{\metre} grating spectrometer. Right: Normalized reflection spectra for \(\mathrm{d_{nom}} = \qty{1800}{\nm}\) and \(\mathrm{x} = 0.10\) measured from \qty{4}{}-\qty{295}{\kelvin} in \qty{50}{\kelvin} steps, showing resonances and the temperature-dependent band edge. Spectra are vertically shifted for clarity.}
\end{figure}

Spectrally resolved reflection measurements are performed in a helium-flow cryostat equipped with an integrated heating stage and motorized lateral translation using mechanical stages. The system enables temperature control from \qty{4}{\kelvin} to \qty{295}{\kelvin}, and spatial positioning over a lateral area of \qtyproduct{25 x 25}{\milli\meter}, cref. Fig.~\ref{fig:setup}. The cryostat is operated under high vacuum conditions at a base pressure of \qty{1e-7}{\milli\bar}.

A commercial broadband thermal emission source (halogen lamp) is focused onto the sample using an achromatic microscope objective of a numerical aperature of \num{0.55}. Reflected light is collected in a backscattering configuration and directed through a \num{5}:\num{95} beamsplitter toward a \qty{0.5}{\meter} long spectrometer equipped with a \qty{300}{lines \per\milli\meter} ruled grating and a Silicon CCD detector sensitve in the \qtyrange{450}{1150}{\nm} spectral range. Reflectance spectra are obtained from normalization of the measured signal to that of a gold reference reflector under identical illumination conditions, effectively compensating for spectral characteristics of the light source and optical components.

For each membrane sample, reflection spectra are recorded at five lateral positions at seven discrete temperatures: \qtylist{4;50;100;150;200;250;295}{\kelvin}, cf. Fig.~\ref{fig:setup}. In parallel, micro-photoluminescence \(\mu \mathrm{PL}\) spectra are obtained under continuous-wave excitation at \qty{532}{\nm}. The excitation laser is coupled into the optical path via a flip mirror (not shown in Fig.~\ref{fig:setup}) and focused onto the same sample as described above.

\subsection{\label{sec:methods.3} Modeling and FDTD Simulations}
The temperature dependence of the bandgap in the \(\mathrm{AlGaAs}\) system has been modeled by Adachi~\textit{et al.}~\cite{adachi1985gaas}, who applied the Varshni relation~\cite{varshni1967temperature} to capture the band edge shift. In this approach, the temperature-dependent bandgap is expressed as a combination of an aluminum-concentration–dependent zero-temperature term and a thermal contribution derived from GaAs. Vurgaftman~\textit{et al.}~\cite{vurgaftman2001band} later refined this description by interpolating the Varshni parameters between GaAs and AlAs. While both formulations yield comparable results over the full composition range, the Vurgaftman model is widely regarded as more complete and is used in this work, as given in Eq.~\ref{eq:Vurgaftman}:

\begin{equation}
\label{eq:Vurgaftman}
E_{g}(x, T) = E_{0}(x, \qty{0}{\kelvin}) - \frac{\alpha(x) \, T^{2}}{T+\beta(x)}
\end{equation}
\[
\begin{aligned}
    &E_{0}(x,\qty{0}{\kelvin}) = 1.519 + 1.155x + 0.37x^{2} \quad \mathrm{(eV)}\\
    &\alpha(x) = \left[5.405(1-x) + 8.85x\right] \cdot 10^{-4} \; \mathrm{(eV/K)}\\
    &\beta(x)  = 204(1-x) + 530x \; \mathrm{(K)}
\end{aligned}
\]

where \(E_{g}\) is the bandgap energy, \(x\) is the aluminum concentration, \(T\) is the temperature, and \(E_0(x, 0\,\mathrm{K})\) is the bandgap at \qty{0}{\kelvin}.\\

Although the Varshni model performs well at elevated temperatures, it tends to overestimate the bandgap narrowing at cryogenic temperatures (\(T < \qty{50}{\kelvin}\)). To address this limitation, Paessler~\textit{et al.}~\cite{passler2003semi} introduced a semiempirical formulation that incorporates phonon freezing effects and provides improved agreement with experimental data in the low temperature regime\cite{lourencco2001temperature}. The model is given by:

\begin{equation}
\label{eq:Paessler}
E_{g}(x,T)=E_{0}(x,\qty{0}{\kelvin})-\frac{\alpha(x) \, \Theta(x)}{2}\left[\sqrt[p(x)]{1+\left(\frac{2 T}{\Theta(x)}\right)^{p(x)}}-1\right]
\end{equation}

\[
\begin{aligned}
    &E_{0}(x,\qty{0}{\kelvin}) = 1.517 + 1.23 x \quad \mathrm{(eV)}\\
    &\alpha(x) = \left(4.9 + 0.7x + 3.7x^2\right) \times 10^{-4} \quad \mathrm{(eV/K)}\\
    &\Theta(x) = 202 + 5x + 260x^2 \quad \mathrm{(K)} \, ,
\end{aligned}
\]

where \(p(x)\) is an empirical (fractional) parameter related to the underlying electron-phonon spectral functions and determined by a non-linear least-squares parameter estimation of the experimental data. It is expected to be larger than a quadratic power dependence \(p > 2\), but the real profile is still under discussion \cite{passler2003semi}.\\

The refractive index of \(\mathrm{AlGaAs}\) can be modeled by a Sellmeier-like dispersion relation first proposed by Adachi~\textit{et al.}~\cite{adachi1985gaas}.  This model provides an analytical form for the dispersive refractive index over composition \(x\) and wavelength \(\lambda\) at RT in the direct bandgap regime \(x < 0.45\) and applies to energies well below the band edge:

\begin{equation}
\label{eq:Adachi}
    n(\lambda, x) = \sqrt{
        A_{0}(x) \left[ f(\chi) + \frac{f(\chi_0)}{2} \left( \frac{E_0}{E_0 + \Delta_0} \right)^{3/2} \right] + B_{0}(x)
    }
\end{equation}

\[
\begin{aligned}
    &f(\chi) = \frac{2 - \sqrt{1 + \chi} - \sqrt{1 - \chi}}{\chi^2}, \\
    &\chi = \frac{hc}{\lambda E_0}, \quad \chi_0 = \frac{hc}{\lambda (E_0 + \Delta_0)}, \\
    &A_0(x) = 6.3 + 19.0x, \quad B_0(x) = 9.4 - 10.2x\\
    &E_0(x) = 1.425 + 1.155x + 0.37x^2 \quad \mathrm{(eV)}\\
    &E_0(x) + \Delta_0(x) = 1.765 + 1.155x + 0.37x^2 \quad \mathrm{(eV)}
\end{aligned}
\]

where \(n\) is the refractive index and \(E_0(x, T)\) the direct bandgap energy at \( T \to \qty{0}{\kelvin}\).\\

To determine the refractive index across the full parameter space defined by the aluminum composition, temperature, and wavelength, we employ an inverse extraction methodology based on resonance-mode matching. Reflection spectra are measured as described in Sec.~\ref{sec:methods.2}, and the spectral positions of optical resonances are extracted from the experimental data in the range between the band edge and \qty{1100}{\nano\meter}. For each Al composition and temperature, measurements are performed at five distinct gold micro cavities on the sample to account for spatial inhomogeneities.

In parallel, finite-difference time-domain (FDTD) simulations are conducted using a commercial solver. The simulated structures replicate the experimental cavity geometry, incorporating the \AlGaAs{} membrane thicknesses (\(\mathrm{d_{eff}}\)) extracted via AFM in Table~\ref{tab:height}. The optical constants used in the simulation include a wavelength dependent refractive index for gold according to Ref.~\cite{johnson1972optical}. For each parameter set, reflection spectra are simulated over a refractive index range of \(n = 3.2\) to \(3.8\). The experimentally observed resonance wavelengths are then compared with the resonances of the simulated spectra. Consequently, the wavelength dependent refractive index is determined by identifying the best-matching simulated set of resonances. As a first step, the temperature-dependent refractive index of \GaAs{} is determined, as it serves as a reference for samples with higher aluminum content (\(x = 0.10, 0.15, 0.30, 0.50)\), all of which feature thin GaAs capping layers that affect the optical response.

To investigate the effect of the developed refractive index model on the optical performance of realistic quantum photonic devices, additional FDTD simulations are conducted for hemispherical microlens structures with and without anti-reflection (ARC) coatings. The simulated geometry consists of a \qty{1}{\micro\meter} gold mirror, a lower \AlGaAs{} membrane of thickness \(\mathrm{d_{QD-m}}\) with refractive index \(n(\mathrm{Al_{nom}}=0.15, T=4 \, K)\) modeled in Sec.~\ref{sec:result.2}, and a quantum emitter modeled as a dipole point-source positioned at the top interface of the membrane. A hemispherical lens of radius \(r = \qty{1350}{\um} + \mathrm{Lens_{v.pos}}\) and \(n(\mathrm{Al_{nom}}=0.15, T=4 \, K)\) is placed centrally atop the dipole source. The geometry is chosen to closely follow our previously demonstrated QD-entangled photon pair sources, further technical details of these FDTD simulations can be found in Ref. ~\cite{langer2025bright}.

\section{\label{sec:results}Results}
\subsection{\label{sec:result.1} Band edge shift and extraction of Al concentration}
\begin{figure}[htbp]
\centering
\includegraphics[width=\linewidth]{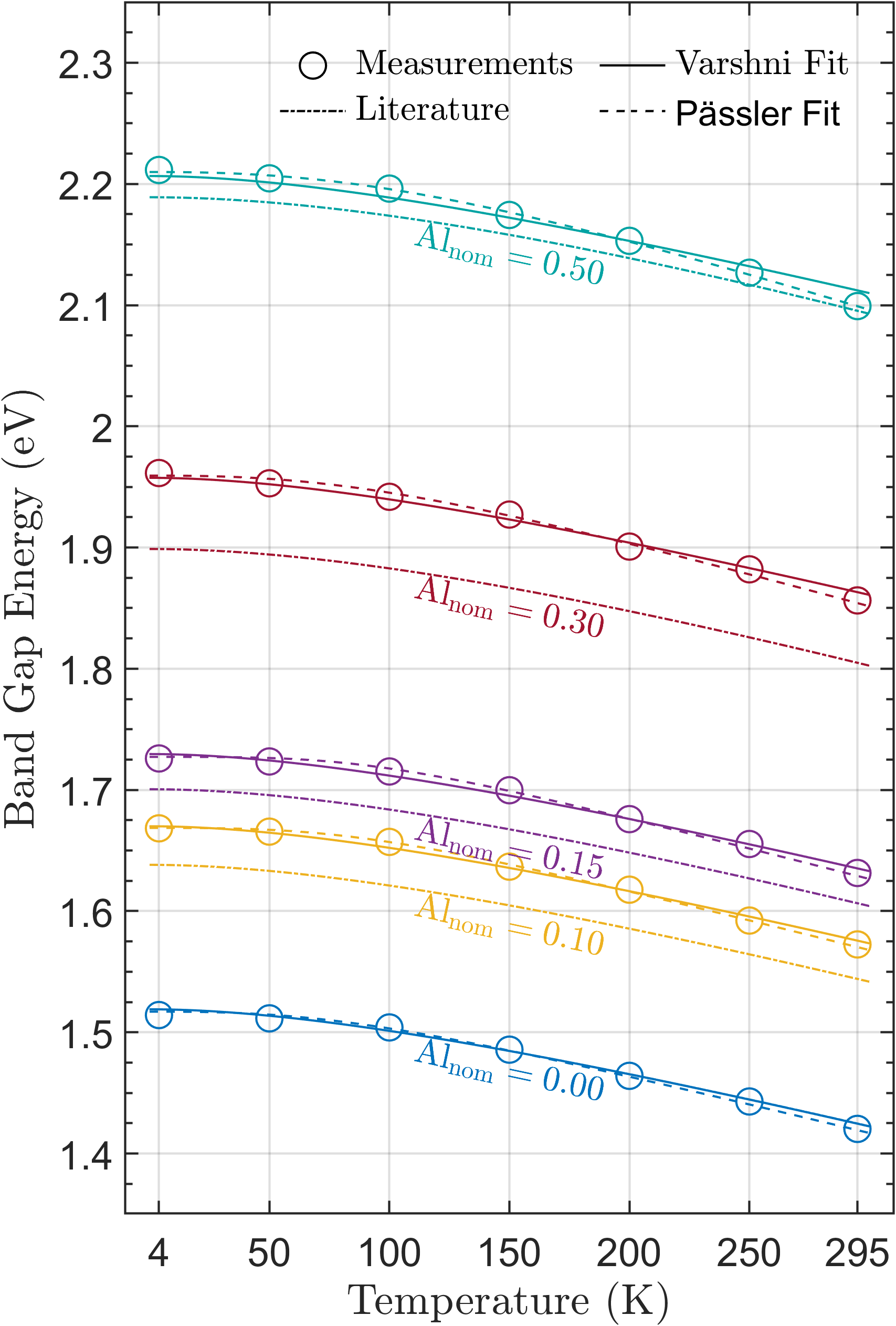}
\caption{\label{fig:Eg_vs_T}
Temperature dependence of the bandgap in \(\mathrm{Al_{x}Ga_{1-x}As}\) for nominal aluminium compositions \(\mathrm{Al_{nom}} = 0.0\,\text{–}\,0.5\). Experimental data are modeled using the Varshni model (Eq.~\ref{eq:Vurgaftman}) and the Paessler model (Eq.~\ref{eq:Paessler}) to extract the band edge and effective aluminium content. Literature reference curves correspond to Eq.~\ref{eq:Vurgaftman} evaluated at the nominal compositions.}
\end{figure}

In addition to variations in membrane thicknesses (Sec.~\ref{sec:methods.1}), fluctuations in aluminum composition can arise during the MBE growth process. To determine the effective Al content, we extract the temperature-dependent bandedge from reflectance spectra using the models of Vurgaftman \textit{et al.}~\cite{vurgaftman2001band} and Paessler \textit{et al.}~\cite{passler2003semi}. Representative modeling to the experimental data is shown in Fig.~\ref{fig:Eg_vs_T}, alongside reference values from Vurgaftman (dash-dotted lines) for the nominal Al compositions \(\mathrm{Al_{nom}}\). The extracted compositions shown in Table~\ref{tab:fit} \(\mathrm{Al_{V}}\) and \(\mathrm{Al_{P}}\) correspond to the Vurgaftman and Paessler models, respectively. The modeling parameter \(p(x)\) is summarized in Appendix~\ref{appendix.1}..

Accurately identifying the absorption onset from reflectance spectra is nontrivial because of the superposition of intrinsic material absorption and photonic resonances. To cross-validate the extracted band edge compositions (\(\mathrm{Al_{V}}\)), we compare them with independent estimates from low- and high-temperature photoluminescence (\(\mathrm{Al_{PL}}\)) and room-temperature X-ray photoelectron spectroscopy (\(\mathrm{Al_{XPS}}\)). The \(\mathrm{Al_{PL}}\) values are systematically lower, consistent with a Stokes shift that cannot be quantitatively determined from the data. As reported in the literature, the magnitude of this shift can vary by several meV depending on factors such as crystal quality and defect density~\cite{ullrich2015inherent}. While PL-based estimates qualitatively support the band edge analysis, they do not permit accurate compositional quantification.

To address this limitation, we employ XPS as a complementary, independent photo-optical characterization method. This surface-sensitive technique enables selective probing by limiting the acceleration voltage to \qty{10}{\kilo\volt} and the depth of electron escape to - at most - the membrane thickness, while the underlying gold mirror additionally acts as an electron drain, suppressing contributions from buried layers. From the resulting spectra, the elemental composition is directly extracted. The values obtained by XPS (\(\mathrm{Al_{X}}\)) show excellent agreement in the direct regime with those derived from the Vurgaftman band edge model (\(\mathrm{Al_{V}}\)), within experimental uncertainties, thus reinforcing the validity of the reflectance-based compositional analysis. XPS underestimates the Al content at \(x = 0.5\), likely due to rapid surface oxidation that takes place in processing and has shifted binding energies not included in the standard analysis. In conclusion, systematic deviations of 2–4\% above the nominal Al content are observed, consistent with known calibration limitations in MBE growth process. In the direct bandgap regime \(x < 0.45\), both models, cf. Eq.~\ref{eq:Vurgaftman} and Eq.~\ref{eq:Paessler}, yield comparable results. The Paessler model, while more accurate at cryogenic temperatures, tends to slightly overestimate the Al content near the indirect transition. Overall, both models offer comparable estimates of the Al content. Given its extensive validation in the literature, we adopt the Varshni-derived values \(\mathrm{Al_{V}}\) as reference for further analysis.

\begin{table}[htbp]
\caption{\label{tab:aluminum} Nominal membrane thickness \(\mathrm{d_{nom}}\) and Al concentration \(\mathrm{Al_{nom}}\) with effective Al concentration by band edge model of Vurgaftman\cite{vurgaftman2001band} (Varshni) \(\mathrm{Al_{V}}\) , band edge model of Paessler\cite{passler2003semi} \(\mathrm{Al_{P}}\), PL spectroscopy formula of Vurgaftman\cite{vurgaftman2001band} \(\mathrm{Al_{PL}}\) and RT XPS \(\mathrm{Al_{X}}\).}
\begin{ruledtabular}
\begin{tabular}{llllll}
\(\mathrm{d_{nom}}\) (nm)
&\(\mathrm{Al_{nom}}\)
&\(\mathrm{Al_{V}}\)
&\(\mathrm{Al_{P}}\)
&\(\mathrm{Al_{PL}}\)\footnote{at \(\mathrm{200\,K}\) and \(\mathrm{4\,K}\)}
&\(\mathrm{Al_{X}}\)\\
\hline
1800 &0.00&0.000&0.000&0.000 / 0.000&0.000\\
1800 &0.10&0.126&0.123&0.123 / 0.120&0.130(5)\\
1460 &0.15&0.173&0.171&0.168 / 0.164&0.170(8)\\
1800 &0.15&0.177&0.176&0.174 / 0.173&0.172(8)\\
1800 &0.30&0.342&0.359&0.335 / 0.339&0.322(8)\\
1800 &0.50&0.511&0.563& - / -  &0.470(9)\\
\end{tabular}
\end{ruledtabular}
\end{table}

\subsection{\label{sec:result.2} Temperature-dependent refractive index \(n(x,\lambda,T)\)}
As a next step, we evaluate the refractive index \(n\) as a function of aluminum composition \(x\), wavelength \(\lambda\), and temperature \(T\). All samples are nominally undoped, and the fabrication process is optimized to minimize residual mechanical stress, ensuring that extrinsic effects on the refractive index are negligible. Furthermore, membrane thicknesses are accurately determined by AFM measurements, and lattice contraction or expansion effects across the temperature range \qtyrange{300}{4}{\kelvin} can be safely neglected~\cite{zhang2020material}.

\begin{figure}
\centering
\includegraphics[width=\linewidth]{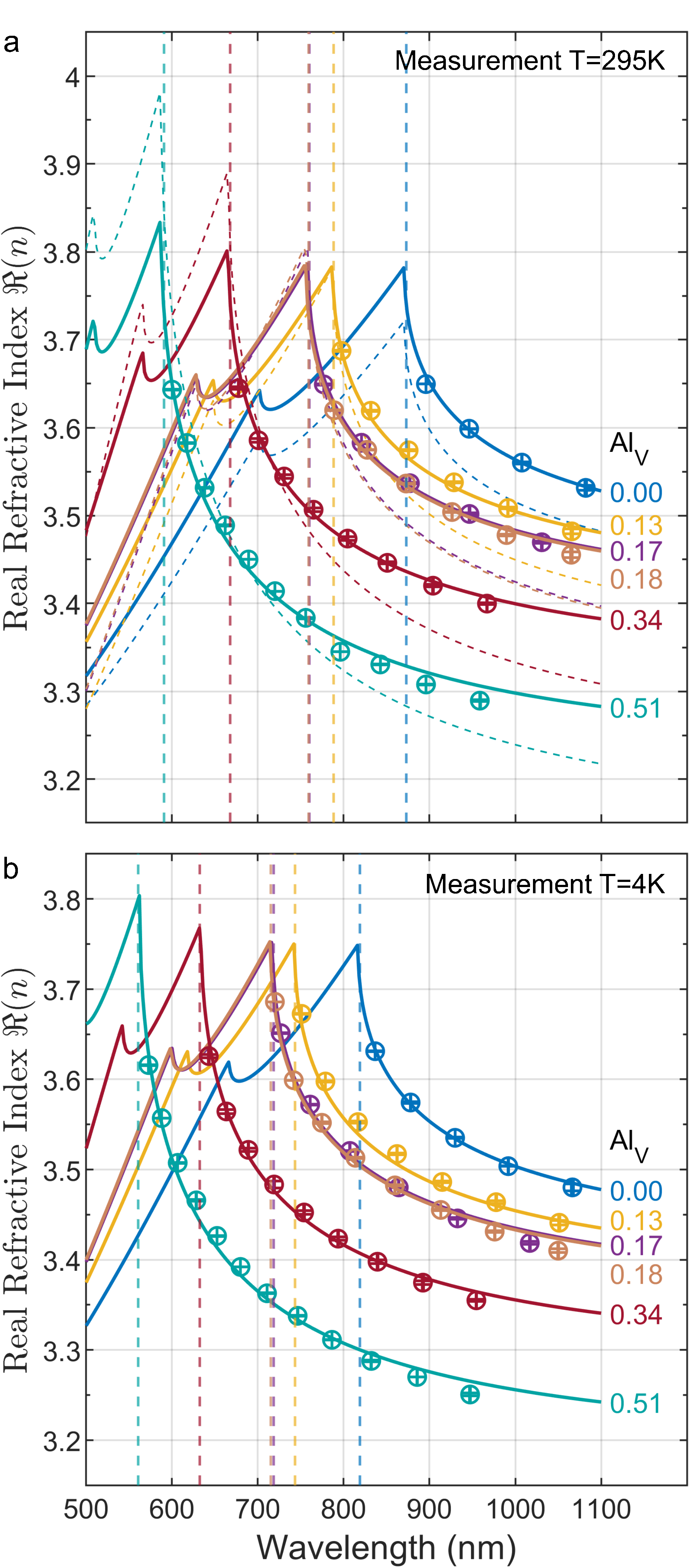}
\caption{\label{fig:ref_index_meas}Refractive index at \qty{295}{\kelvin} and \qty{4}{\kelvin} extracted from experimental reflectance spectroscopy (dots with error bars). The dashed lines show the modelling according to Adachi \textit{et al.}~\cite{adachi1985gaas} as given in Eq.~(\ref{eq:Adachi}), while solid lines represent the refined Model 3 introduced in this work, Eq.~(\ref{eq:Ours}).}
\end{figure}

On the basis of these assumptions, we extract the dispersive refractive index for each sample and temperature by remodeling the experimental resonances from reflectance spectroscopy using FDTD simulations. The precise structural parameters, detailed in Sec.~\ref{sec:methods.1} and Sec.~\ref{sec:result.1}, are used as simulation inputs, with the spectral output yielding theoretical resonances within a refractive index range of \qtyrange{3.2}{3.8}{}. The experimental resonance positions are matched to the simulated spectra, enabling the extraction of the best parameter estimate values \(n\). The method is illustrated and explained in great detail in Appendix~\ref{appendix.5}.

Figure~\ref{fig:ref_index_meas} shows the extracted refractive indices at both RT and \qty{4}{\kelvin}, compared to the values in the literature. We first consider the standard model by Adachi \textit{et al.}~\cite{adachi1985gaas} (Model~0), incorporating a temperature-dependent band edge shift via Eq.~\ref{eq:Vurgaftman}. This serves as the baseline for comparison against a series of increasingly refined models introduced in this work. Model~1 retains the same functional form as Model~0, but allows the dispersion parameters \(A_0(x)\) and \(B_0(x)\) to vary quadratically with composition. Model~2 further incorporates a first-order temperature-dependent term \(C_0(x)\), and Model~3 includes an additional second-order temperature term \(D_0(x)\), resulting in a full expression of the form in Eq.~\ref{eq:Ours}.

To assess the performance of the model, we use the root mean square error (RMSE) and the coefficient of determination (\(R^2\)), averaged over all compositions of aluminum and temperatures. To evaluate model compactness and avoid overparameteriazation, we additionally consider the adjusted \(R^2\), as discussed in detail in Appendix~\ref{appendix.4}.

Model~0 shows reasonable agreement with experimental data near the absorption edge across all temperatures but systematically underestimates the refractive index at longer wavelengths, with deviations up to \(\Delta n \approx 0.1\). It yields an RMSE of 0.0385 and \(R^2 = 0.8577\). Moreover, a critical reassessment of the original data by Adachi \textit{et al.}~\cite{adachi1985gaas} suggests that even their own measurements are slightly underestimated by their model, presumably due to limitations in the model itself.

By remodeling \(A_0(x)\) and \(B_0(x)\) as second-order polynomials (Model~1), the predictive accuracy improves substantially, reducing the RMSE by nearly 50\% and increasing \(R^2\) by +0.1057. However, Model~1 still performs better at low temperatures than at room temperature. This discrepancy is further mitigated in Model~2, which introduces a linear temperature term \(C_0(x)\), reducing the RMSE to 0.0097 and achieving \(R^2 = 0.9909\), representing a fourfold improvement in RMSE over Model~0.

Finally, inclusion of the second-order temperature term \(D_0(x)\) in Model~3 yields the best overall performance, with RMSE = 0.0083 and \(R^2 = 0.9934\). This confirms that the proposed model reliably predicts the refractive index across the full wavelength, composition, and temperature space, while maintaining a compact functional form and avoiding overparameteriazation. All model parameters are physically motivated and statistically justified, as discussed in Appendix~\ref{appendix.4}.

\begin{figure}
\centering
\includegraphics[width=\linewidth]{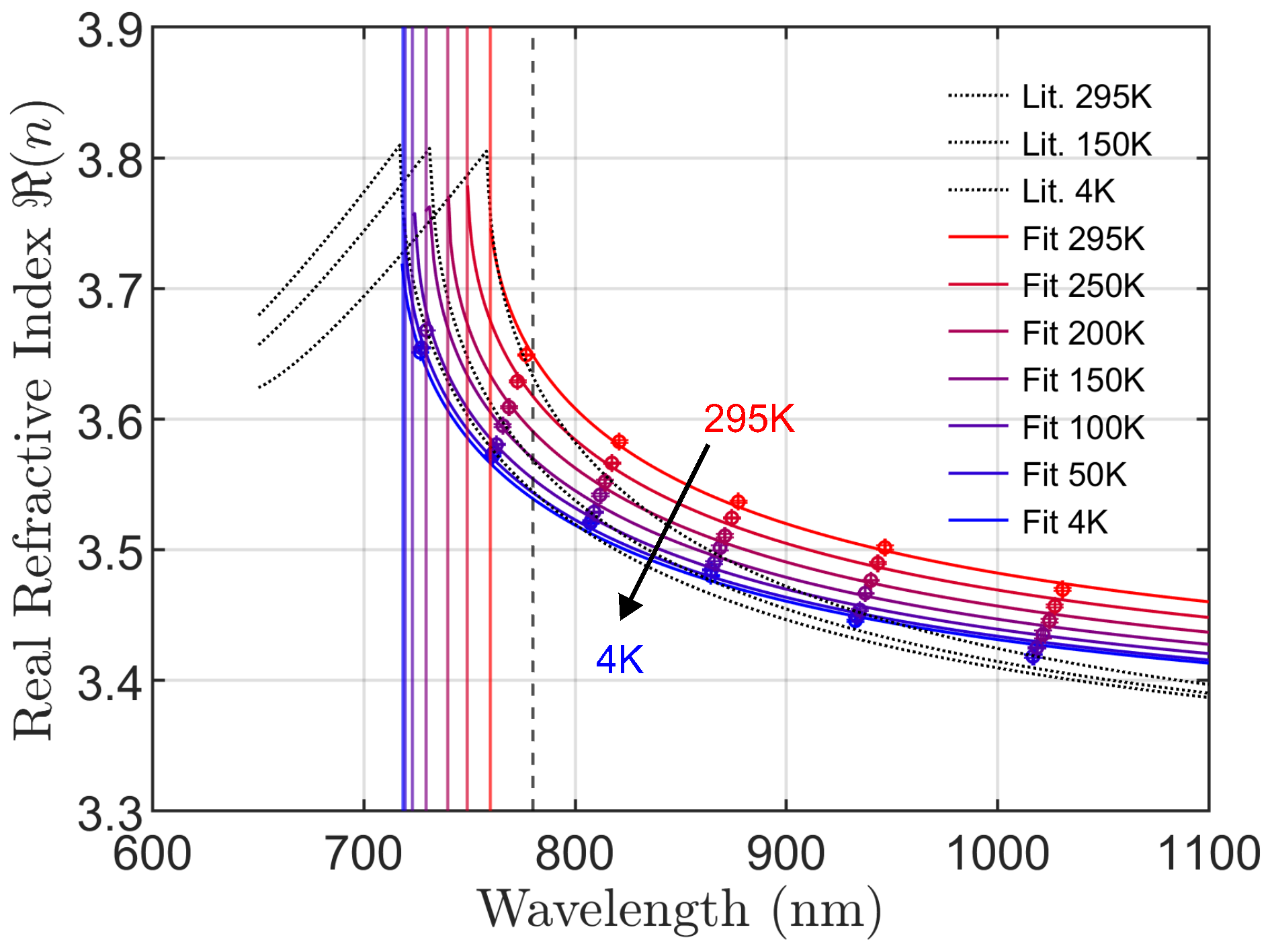}
\caption{\label{fig:T_series}
Temperature-dependent refractive index of \AlGaAs{} with \(\mathrm{Al}_\mathrm{V} = 0.171\), measured from \qty{295}{\kelvin} to \qty{4}{\kelvin}. Experimental data are compared to the Adachi model (black dotted lines), including a band edge shift correction~\cite{adachi1985gaas}. Solid lines show the proposed model across all temperatures, from room temperature (red) to cryogenic temperatures (blue), demonstrating excellent agreement.}
\end{figure}

The final expression for the refractive index is given by:

\begin{equation}
\label{eq:Ours}
    n(\lambda, x, T) = n(\lambda, x) + C_0(x) T + D_0(x) T^2 \,
\end{equation}

\[
\begin{aligned}
    A_0'(x) &= 6.741 + 2.938x + 11.686x^2 \, ,\\
    B_0'(x) &= 9.275 - 2.489x - 6.940x^2 \, ,\\
    C_0(x) &= -2.618 \times 10^{-6} \, \mathrm{K}^{-1},\\
    D_0(x) &= 4.282 \times 10^{-7} \, \mathrm{K}^{-2}.
\end{aligned}
\]
Where \(n(\lambda, x)\) is given by Eq. \ref{eq:Adachi}.
\\
A direct comparison between Model~0 and Model~3 at a fixed aluminum concentration of \(\mathrm{Al}_\mathrm{V} = 0.171\) is shown in Fig.~\ref{fig:T_series}, illustrating the incorporation of the temperature-dependent band edge shift into the Adachi model~\cite{adachi1985gaas}. Although Model~0 accounts for this shift, it fails to accurately reproduce the spacing between refractive index curves at different temperatures, particularly at longer wavelengths. The intermediate remodeling in Model~1 improves this aspect, while Model~3 achieves excellent agreement with the experimental data across the entire temperature range. This comparison highlights the pivotal role of the band edge shift in determining the dispersive behavior of the refractive index. The strong correlation between its temperature dependence and the evolution of \(n\) confirms the band edge shift as the primary driver of thermal refractive index variation. The parameters \(A_0(x)-D_0(x)\) are employed to account for the nature of this correlation.

\subsection{\label{sec:result.3} Application in quantum photonics}

The continued development of \GaAs{} quantum dots (QDs) as ultra-compact~\cite{langer2025ultra} and maximally entangled~\cite{hopfmann2021maximally} single-photon sources increasingly requires precise knowledge of the refractive index as a critical design parameter in photonic device engineering. As demonstrated by the structural thickness deviations in Table~\ref{tab:height} and the aluminum concentration variations in Table~\ref{tab:aluminum}, refractive index values can vary due to inherent fabrication imperfections. To benchmark these effects, a reference sample with a nominal thickness of \(d_{\mathrm{nom}} = \qty{1460}{\nano\meter}\) was fabricated and analyzed alongside the microcavity series. This sample exhibited excellent agreement with the refractive index trends extracted from the broader data set (cf. Fig.~\ref{fig:ref_index_meas}), seamlessly integrating into the established model. This consistency underscores both the reliability and the general applicability of the extraction methodology presented across a diverse set of microcavity geometries. Moreover, the closed-form model enables inverse parameter estimation, allowing post-fabrication identification of membrane thickness, operating temperature, or aluminum concentration, offering a robust route to compensate for fabrication-induced deviations.

\begin{figure}
\centering
\includegraphics[width=\linewidth]{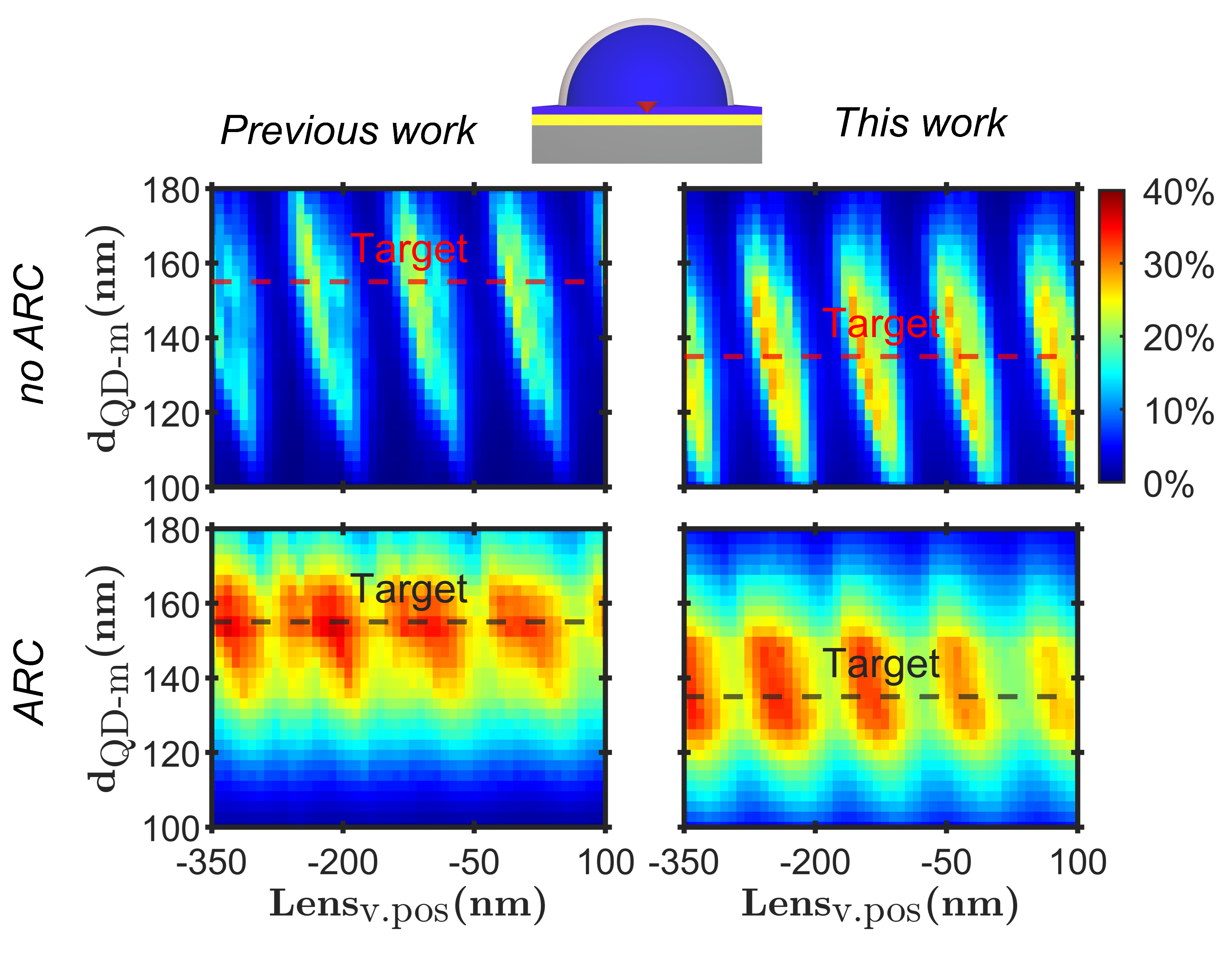}
\caption{\label{fig:Comparison_for_Membrane_Paper}
FDTD simulation of the fiber-coupled emission efficiency \(\eta_{\mathrm{FE}}\) for a hemispherical microlens integrated onto a gold mirror, as schematically shown at the top. The structure consists of a quantum dot located at the center of a microlens with radius defined as \(r = \qty{1350}{\nano\meter} + \mathrm{Lens_{v.pos}~(nm)}\), and a vertical separation \(d_{\mathrm{QD-m}}\) between the quantum dot and the gold mirror. The design includes an ARC of thickness \qty{112}{\nano\meter} and refractive index \(n = 1.54\). Colored target lines denote previously reported optimal QD position and cavity lengths for maximum coupling efficiency to a 780nm HP single-mode fiber.}
\end{figure}

To illustrate the practical implications of the refined refractive index model, we compare it with values used in recent quantum photonic device studies. For example, Chen \textit{ et al.}~\cite{chen2018highly} employed \(n = 3.500\), while Nie \textit{et al.}~\cite{nie2021experimental} and Langer \textit{et al.}~\cite{langer2025bright} used \(n = 3.312\), and Bremer \textit{et al.}~\cite{bremer2022numerical} adopted \(n = 3.347\), all evaluated at \(\lambda = \qty{780}{\nano\meter}\) and \(T = \qty{4}{\kelvin}\). According to our experimentally validated model, see Fig.~\ref{fig:ref_index_meas}, the refractive index under these conditions is \(n = 3.561\), indicating a systematic deviation from previous assumptions.

Figure~\ref{fig:Comparison_for_Membrane_Paper} quantifies the impact of this updated refractive index on fiber-coupled emission efficiency \(\eta_{\mathrm{FE}}\) for quantum dots embedded in hemispherical microlens structures on a gold-mirror. Two key design parameters are varied: the quantum dot to gold-mirror distance \(d_{\mathrm{QD-m}}\) and the microlens radius\,\(r\). This geometry has previously been studied~\cite{langer2025bright}, and is here refined using an improved optical simulation relying on the refined refractive index estimation.

Earlier simulations identified optimal emission efficiency at \(d_{\mathrm{QD-m}} = \qty{155}{\nano\meter}\), both with and without ARC, as indicated by the red and black dotted lines. With the updated refractive index model, we observe that these previous target distances only marginally align with the optimal parameter ranges. The revised model shifts the predicted optimum to \(d_{\mathrm{QD-m}} \approx \qty{135}{\nano\meter}\). Furthermore, the resonance spacing for the microlens radius is reduced, which is consistent with a higher effective refractive index. Importantly, while the absolute fiber-coupled emission efficiency remains largely unchanged, the refined model significantly alters the optimal geometrical parameters required to achieve it.

\section{Conclusions}

In summary, we present a simplified and accurate parameterization method for the complex refractive index of \(\mathrm{Al_{x}Ga_{1-x}As}\) in the direct bandgap regime (\(x = 0\,\text{-}\,0.5\)) and the temperature range (\(T = 295\,\text{-}\,\qty{4}{\kelvin}\)) for wavelengths from the bandgap up to \qty{1100}{\nano\meter}. The resulting analytical expression enables the evaluation of both the refractive index \(n\) and the extinction coefficient \(k\) over the entire parameter space, offering practical improved estimation values with minimal deviation from experimental observations with an average error as small as RMSE=0.0083 and variance predictability of \(R^2=0.9934\). Beyond its modeling capabilities, we demonstrated the direct applicability of this formalism to integrated quantum photonics. The method not only serves as a reliable control tool for assessing fabrication accuracy, such as membrane thickness, temperature, or aluminum content, but also contributes to the optimization of photonic structures operating near the band edge, where small deviations in refractive index significantly affect device performance. This work lays the foundation for more precise and fabrication-aware photonic device engineering, bridging the gap between material science and quantum photonics.

\begin{acknowledgments}
We thank the clean room team, especially Ronny Engelhard and Martin Bauer, of the Leibniz IFW Dresden for his efforts and expertise in clean room processing of samples and preparation of scanning electron beam and focused ion beam images. We would also like to thank G. Botzen for the fruitful and inspiring discussions. This work was funded by the German Federal Ministry of Education and Research (BMBF) projects QR.N, QUARKS, QUIET and QD-CamNetz (contracts no. 16KIS2194, 16KIS1998K, 16KISQ094, and 16KISQ078).
\end{acknowledgments}

\section*{Data availability statement}
The full dataset that supports the findings presented in this study is available upon reasonable request from the corresponding author.

\appendix
\section{\label{appendix.1}Experimental band edge modeling}

The empirical parameter \(p(x)\) in Eq.~\ref{eq:Paessler} extends the classical Varshni model by incorporating electron–phonon interactions and accounting for phonon dispersion effects~\cite{passler2003semi}. While \(p(x)\) has been widely used to describe temperature-dependent band edge shifts in various binary and ternary III–V semiconductors, it remains an experimentally determined parameter. Lourenço \textit{et al.}~\cite{lourencco2001temperature} demonstrated its extraction through temperature-resolved optical spectroscopy, but a predictive theoretical model remains elusive. Table~\ref{tab:fit} compares the experimentally extracted values from this work (see Sec.~\ref{sec:result.1}) with those reported in the literature showing a congruent agreement for small aluminum concentrations.

\begin{table}[H]
\caption{\label{tab:fit} 
Comparison of the experimental modeling parameter \(p(x)\) from Paessler's model~\cite{passler2003semi} extracted for our experimental results with values reported by Lourenço \textit{et al.}~\cite{lourencco2001temperature}. The nominal aluminum concentration of 0.15 used in this work and 0.17 in the literature are considered equivalent for this comparison.}
\begin{ruledtabular}
\begin{tabular}{lccccccc}
\(\mathrm{Al_{nom}}\) & 0.00 & 0.10 & 0.15/0.17 & 0.30 & 0.40 & 0.50 \\
\hline
\(p(x)\) (this work) & 2.75 & 3.19 & 3.69 & 2.56 & -- & 2.44 \\
\(p(x)\) (Ref.~\cite{lourencco2001temperature}) & 2.85 & -- & 4.00 & 3.50 & 2.90 & -- \\
\end{tabular}
\end{ruledtabular}
\end{table}

\section{\label{appendix.5} Modeling and refractive index extraction}

\begin{figure} [H]
\centering
\includegraphics[width=\linewidth]{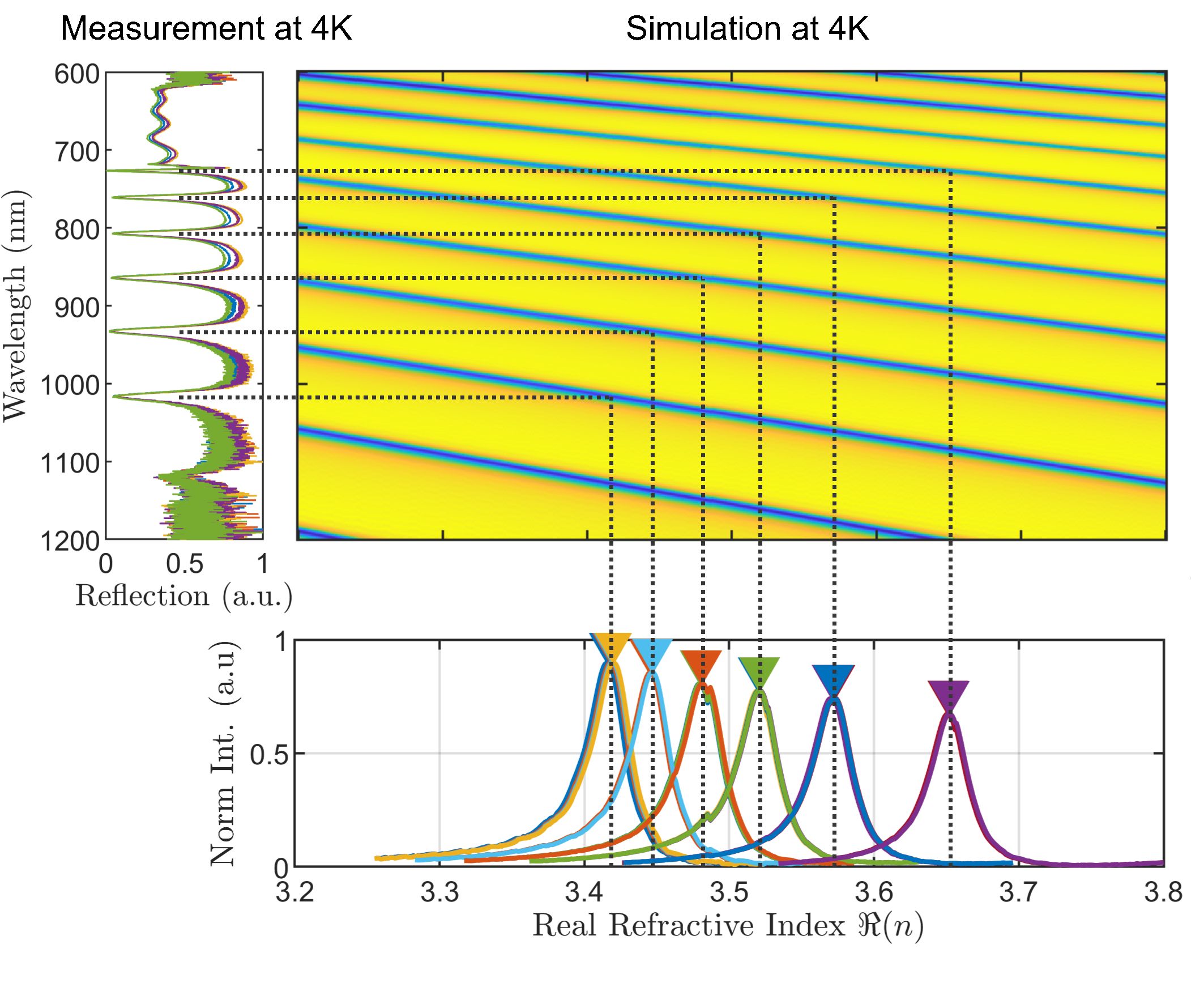}
\caption{\label{fig:simulation_expl}
Determination of the refractive index from reflectance spectroscopy. Using finite-difference time-domain simulations (center) of the membrane cavity structures, the resonant features observed in the reflectance spectra (left) are mapped to their corresponding refractive indices (bottom). The dotted-lines serve as a guidance to the eye.}
\end{figure}

At the center of this publication is the extraction of the refractive index from reflection measurements performed on \AlGaAs{} over a temperature range of \qtyrange{4}{300}{\kelvin}. The workflow of this extraction is illustrated in Fig.~\ref{fig:simulation_expl} for the representative sample with \(\mathrm{d_{nom}}=\qty{1460}{\nm}\) and \(\mathrm{Al_{nom}}=0.15\) at \qty{4}{\kelvin}. The left panel shows the experimentally recorded reflection spectrum between \qtyrange{600}{1200}{\nm}. The pronounced minima in reflectance correspond to regions of enhanced optical absorption within the membrane cavity, which is formed by the \AlGaAs{} membrane sandwiched between a bottom gold mirror and a semi-transparent top gold layer.

The central panel displays the corresponding FDTD simulation using the experimentally determined structural parameters. The \AlGaAs{} membrane thickness is given by \(\mathrm{d_{eff}}\), which includes additional \qty{2}{\nm} (\qty{10}{\nm}) \GaAs{} layers at the top and bottom interfaces for \(\mathrm{Al_{nom}} = 0.10\;\text{-}\,0.30\) (\(\mathrm{Al_{nom}} = 0.50\)), while the gold layers have thicknesses of \qty{100}{\nm} (bottom) and \qty{10}{\nm} (top). The refractive index of these thin \GaAs{} layers was calculated beforehand for all temperatures and subsequently used as fixed input in the simulations for \(\mathrm{Al_{nom}} = 0.10\;\text{-}\,0.50\). The FDTD simulations yield the expected wavelength-dependent reflectance (absorption) spectra for the typical refractive-index range of \AlGaAs{}, i.e. \qtyrange{3.2}{3.8}{}.

The reflection minima observed experimentally correspond to cavity modes (resonances) evolving in the membrane structure. These modes appear in ascending order and can be assigned via the simulation onto a corresponding refractive-index value, as illustrated in the lower panel of Fig.~\ref{fig:simulation_expl}. The procedure is repeated for all five recorded reflection spectra, and the resulting \(n\)-values are averaged to obtain a mean refractive index and its standard deviation. Each projected refractive-index value represents one data point in Fig.~\ref{fig:ref_index_meas}. Repeating this method for all aluminium concentrations and temperatures yields the full dataset \(n(\lambda, x, T)\), which can be plotted either at fixed temperatures (Fig.~\ref{fig:ref_index_meas}) or fixed aluminium concentrations (Fig.~\ref{fig:T_series}).

This method provides an efficient and robust approach to determine the dispersive refractive index while still allowing independent evaluation of the temperature and aluminium concentration dependencies. However, the analysis reveals that an accurate determination of the membrane thickness is crucial: deviations of approximately \qty{30}{\nm} in thickness lead to changes in \(n\) of up to \qty{0.05}{}. Despite the high precision of MBE growth, a reliable post-growth thickness characterization is therefore indispensable. This justifies the labor-intensive mask definition, etching, and AFM measurements required to precisely determine \(\mathrm{d_{eff}}\). Previous studies often relied solely on nominal growth thicknesses, which can deviate from the effective optical thickness, particularly in as-grown structures where the \AlGaAs{} layer is embedded within the monolithic epitaxial crystal and the zero-level not easy to access. In contrast, the membrane-release process employed here defines a clear and accurate reference plane at the gold mirror, offering a significantly improved basis for precise thickness extraction despite the added fabrication complexity.

\section{\label{appendix.4} Parameter evaluation}

In this section, we evaluate the necessity and validity of introducing additional parameters to the modeling of the experimental data. Although it is well known that increasing the number of modeling parameters generally improves a model’s ability to capture the data, overparameterization must be avoided. The inclusion of non-essential parameters does not necessarily yield deeper physical insight and can instead lead to unnecessarily complex expressions. To assess the impact of added complexity, we focus on two widely used statistical metrics: the Root Mean Squared Error (RMSE) and the Coefficient of Determination (\(R^2\)).

The RMSE quantifies the average deviation between observed and predicted values, providing a direct measure of a model's predictive accuracy.

\begin{equation}
\label{eq:RMSE}
    \mathrm{RMSE} = \sqrt{ \frac{1}{n} \sum_{i=1}^{n} \left( y_i - \hat{y}_i \right)^2 },
\end{equation}

where \(y_i\) and \(\hat{y}_i\) denote the measured and predicted values, respectively, and \(n\) is the number of data points. Improvements in RMSE are commonly reported as relative changes, with reductions of \(\geq 10\%\) considered significant and \(\geq 5\%\) generally considered justifiable.

To complement this metric, we evaluate the Coefficient of Determination, \(R^2\), which estimates the proportion of variance in the dependent variable that is captured by the model:

\begin{equation}
\label{eq:R2}
    R^2 = 1 - \frac{SS_{\mathrm{res}}}{SS_{\mathrm{tot}}} = 1 - \frac{ \sum_{i=1}^n \left( y_i - \hat{y}_i \right)^2 }{ \sum_{i=1}^n \left( y_i - \bar{y} \right)^2 },
\end{equation}

where \(SS_{\mathrm{res}}\) and \(SS_{\mathrm{tot}}\) denote the residual and total sum of squares, respectively, and \(\bar{y}\) is the mean of the measured values. An increase in \(R^2\) by \(\geq 0.05\) is generally interpreted as a significant improvement, while increases of \(\geq 0.02\) provide a justifiable model refinement.

To further assess the relevance of additional parameters and avoid overparameterization, we employ the Adjusted Coefficient of Determination (\(Adj. R^2\)):

\begin{equation}
\label{eq:adjR2}
    \mathrm{Adj.} \ R^2 = 1 - \left( \frac{ (1 - R^2)(n - 1) }{ n - p - 1 } \right),
\end{equation}

where \(p\) is the number of model parameters. Unlike \(R^2\), the adjusted version penalizes excessive model complexity. If both RMSE and \(R^2\) improve with the addition of parameters but the adjusted \(R^2\) decreases, the added complexity is not considered justified.\\

Our parameter extension is based on the original formulation by Adachi \textit{et al.} (Eq.~\ref{eq:Adachi}), which does not account for temperature-dependent effects related to the band edge shift. However, this band edge shift will be included in to model and serves as the baseline for our analysis and is hereafter referred to as Model 0. To systematically investigate the impact of increasing model complexity, we introduce a series of parameterizations with progressively more degrees of freedom.

In Model~1, we replace the first-order polynomial expressions for the parameters \(A_0(x)\) and \(B_0(x)\), cf. Eq. \ref{eq:Adachi}, used in Model~0 with second-order polynomials. This modification is physically justified, given the observed curvature in the compositional dependence, and leads to a substantial improvement in model performance by a \qty{33.3}{\percent} reduction in RMSE and a 0.037 increase in \(R^2\) for \(A_0(x)\), and a \qty{45.4}{\percent} reduction in RMSE alongside a 0.031 increase in \(R^2\) for \(B_0(x)\). All subsequent models adopt this second-order representation as a new baseline.

Model~2 extends this framework by introducing an additional first-order temperature-dependent term \(C_0(x)\), capturing a linear change in refractive index with temperature. Finally, Model~3 incorporates a second-order temperature term \(D_0(x)\), approaching the practical limits of parametrization before the onset of overparameterization.

\begin{table*}[htbp]
\caption{\label{tab:parameters} Summary of model parametrizations, model coefficient expressions, and statistical performance metrics for the dispersive composition- and temperature-dependent refractive index of \AlGaAs{}. Model complexity increases from left to right. With Root Mean Square Error RMSE, Coefficient of determination \(R^2\) and Adjusted Coefficient of Determination \(Adj. R^2\).}
\begin{ruledtabular}
\begin{tabular}{lllll}
Model
&0 (Lit.)
&1
&2
&3\\
\hline
Parameters& (A+B)\footnote{first-order} 
          & (A+B)\footnote{second-order} 
          & (A+B)$^{\text{b}}$+C 
          & (A+B)$^{\text{b}}$+C+D \\
Values
&$A_0(x)=6.3+19.0x$
&$A_0(x)=6.731+2.916x+11.879x^2$
&$A_0(x)=6.717+3.146x+11.232x^2$
&$A_0(x)=6.741+2.938x+11.686x^2$\\
&$B_0(x)=9.4-10.2x$
&$B_0(x)=9.361-2.435x-7.090x^2$
&$B_0(x)=9.242-2.514x-6.847x^2$
&$B_0(x)=9.275-2.489x-6.940x^2$\\
&&&$C_0=1.238\times10^{-4}\,\mathrm{K}^{-1}$&$C_0=-2.618\times10^{-6}\,\mathrm{K}^{-1}$\\
&&&&$D_0=4.282\times10^{-7}\,\mathrm{K}^{-2}$\\
RMSE&0.0385&0.0195&0.0097&0.0083\\
\(R^2\)&0.8577&0.9634&0.9909&0.9934\\
Adj.\(R^2\)&0.8560&0.9630&0.9908&0.9933\\
\end{tabular}
\end{ruledtabular}
\end{table*}

An overall comparison of model performance is summarized in Table~\ref{tab:parameters}. The baseline model from the literature (Model~0) achieves an \(R^2\) of 0.857, capturing the general trends of the experimental data. Introducing a second-order polynomial parameterization for \(A_0(x)\) and \(B_0(x)\) in Model~1 yields a substantial improvement, increasing \(R^2\) by +0.1057 to 0.9634, while simultaneously halving the RMSE. Notably, the enhancement is particularly pronounced at low temperatures (\qty{4}{\kelvin}), where an additional \(\Delta R^2 = +0.005\) is observed relative to RT (\qty{295}{\kelvin}).

This temperature-dependent discrepancy is further mitigated in Model~2 through the inclusion of a first-order temperature coefficient \(C_0(x)\). Compared to Model~1, Model~2 exhibits a further increase of \(\Delta R^2 = +0.0275\), with the RMSE again reduced by approximately 50\%.

Extending the model to include a second-order temperature term \(D_0(x)\) in Model~3 results in an additional 14\% reduction in RMSE compared to Model~2. However, the improvement in \(R^2\) is marginal (\(\Delta R^2 = +0.0025\)), indicating diminishing returns. Despite this, the higher Adjusted \(R^2\) still justifies the inclusion of \(D_0(x)\), though further parameterization beyond this point is not warranted.

Through this stepwise model refinement, by introducing second-order composition dependence in \(A_0(x)\) and \(B_0(x)\), and augmenting with temperature-dependent terms \(C_0(x)\) and \(D_0(x)\), we achieve a substantial improvement in model fidelity. Overall, the RMSE is reduced by more than 75\%, and the coefficient of determination increases by +0.1332, reaching an exceptional value of \(R^2 = 0.9934\).

\section{\label{appendix.3}Temperature dependence of \(n\): first- and second-order contributions}

In this section, we examine the influence of the first- and second-order temperature-dependent parameters, \(C_0(x)\) and \(D_0(x)\), on the refractive index \(n\), as introduced in Eq.~\ref{eq:Ours}. These parameters account for the dispersive character of the temperature response and are most apparent in spectral regions where the temperature-induced shift of the band edge is minimal or approximately linear.

While the dominant contribution to the temperature dependence of \(n\) arises from the shift in the absorption edge, \(C_0(x)\) and \(D_0(x)\) introduce spectrally resolved corrections that become especially evident at cryogenic (\qtyrange{0}{50}{\kelvin}) and elevated (\qtyrange{250}{295}{\kelvin}) temperatures. In these regimes, the band edge remains either nearly static or shifts linearly with temperature, allowing the first- and second-order terms to manifest clearly. In contrast, the intermediate range (\qtyrange{50}{250}{\kelvin}) exhibits strong nonlinearity, which obscures the subtle effects of these parameters and is thus excluded from the analysis.

\begin{figure} [H]
\centering
\includegraphics[width=\linewidth]{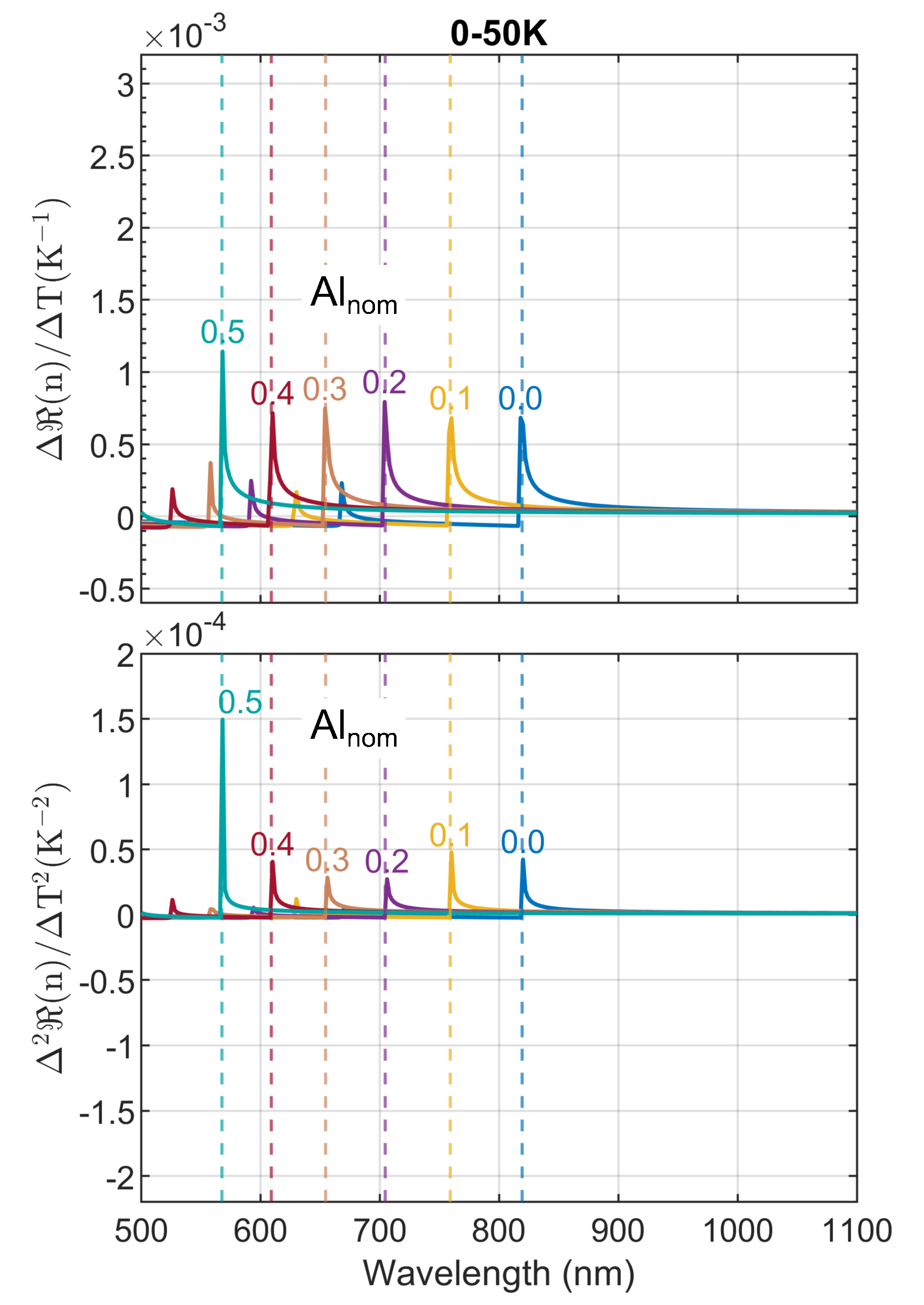}
\caption{\label{fig:050K_dndT_d2ndT2}
Spectral dependence of the first- and second-order temperature derivatives of the refractive index \(n\) for \(\mathrm{Al_{x}Ga_{1-x}As}\) with \(x = 0.0\,\text{–}\,0.5\), evaluated in the cryogenic regime (\(T = 0\,\text{–}\,50\,\mathrm{K}\)). In this range, the band edge remains nearly static, allowing both \(\Delta n/\Delta T\) and \(\Delta^2n/\Delta T^2\) to be resolved clearly. The second-order component exhibits a smooth profile and is physically interpretable, in contrast to the high-temperature behavior.}
\end{figure}

\begin{figure} [H]
\centering
\includegraphics[width=\linewidth]{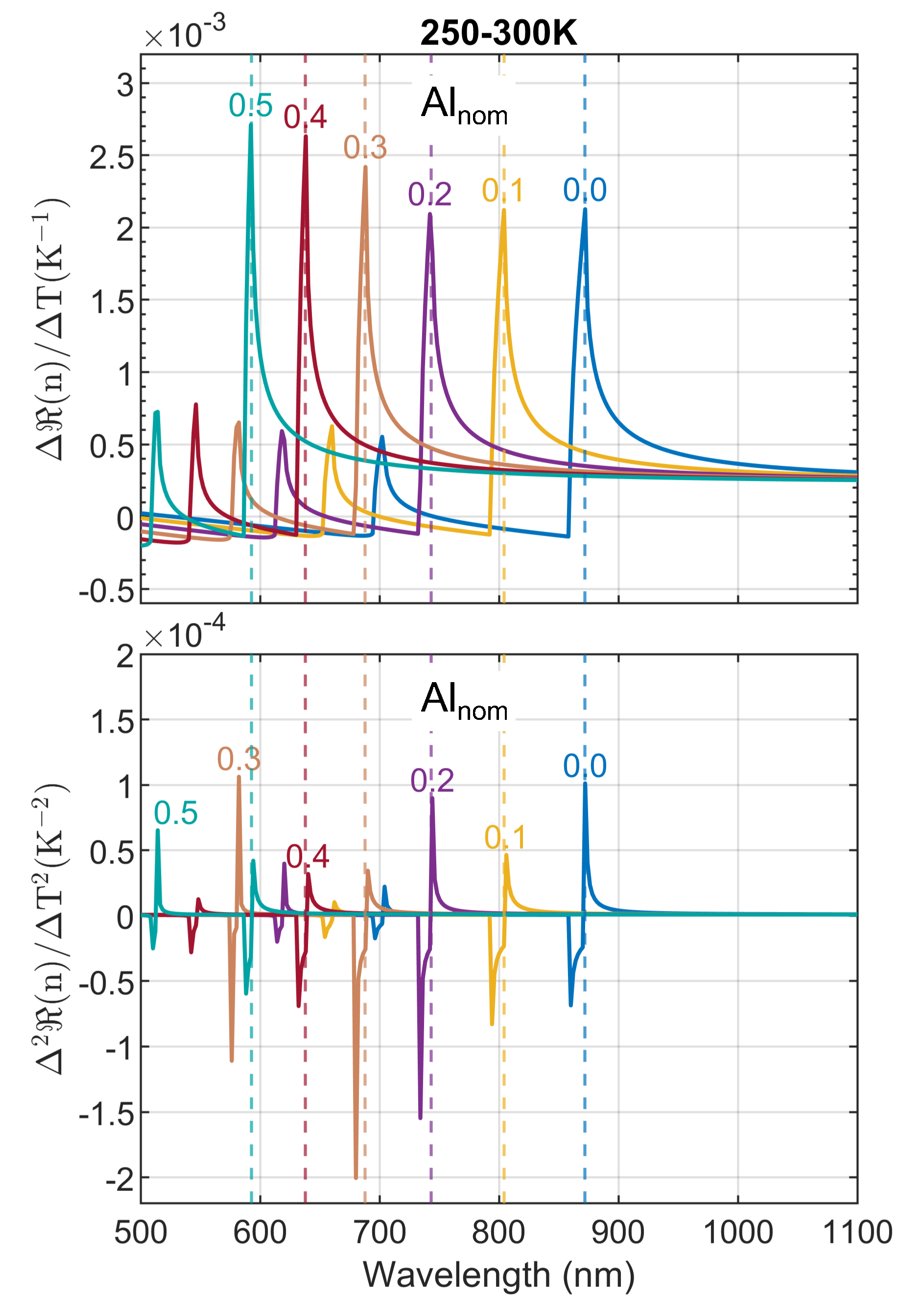}
\caption{\label{fig:250_300K_dndT_d2ndT2}
Spectral dependence of the first- and second-order temperature derivatives of the refractive index \(n\) for \(\mathrm{Al_{x}Ga_{1-x}As}\) with \(x = \mathrm{Al_{nom}} = 0.0\,\text{–}\,0.5\), evaluated in the high-temperature regime (\(T = 250\,\text{–}\,300\,\mathrm{K}\)). The first-order term (\(\Delta n/\Delta T\)) shows a strong peak near the absorption edge, decaying exponentially toward longer wavelengths, while the second-order term (\(\Delta^2n/\Delta T^2\)) oscillates around zero with limited interpretation at elevated temperatures.}
\end{figure}

To quantify these contributions, the first- and second-order derivatives \(\Delta n / \Delta T\) and \(\Delta^2 n / \Delta T^2\) were evaluated within the two linear regimes and averaged. As shown in Fig.~\ref{fig:050K_dndT_d2ndT2} and Fig.~\ref{fig:250_300K_dndT_d2ndT2}, the first-order term governed by \(C_0(x)\) dominates at higher temperatures, where its magnitude increases by a factor of 2–3 relative to cryogenic conditions. The derivative peaks sharply near the band edge and decays exponentially at longer wavelengths. The associated steep gradients, while challenging for experimental validation, are accurately captured in our model. The second-order term, \(D_0(x)\), appears to play a more defined role at low temperatures. Its effect is less but still observable and interpretable in our model. However, at elevated temperatures, it is approximately zero and shows no systematic trend. This conclusion is supported by global modeling metrics shown inf Appendix~\ref{appendix.4}, where inclusion of \(D_0(x)\) yields only marginal improvements and can lead to overparameterization if not constrained.

These findings highlight the critical role of \(C_0(x)\) in enhancing the accuracy of the model across the entire temperature range, with pronounced effects near the absorption edge, where the thermal sensitivity is greatest. In contrast, the influence of \(D_0(x)\) is largely restricted to cryogenic conditions and warrants careful consideration to avoid overparameterization. Together, these parameters enable a more accurate and physically grounded description of refractive index variations with temperature, which is essential for the design and operation of quantum photonic devices requiring high-precision tuning. While thermal modulation at cryogenic temperatures results in minuscule changes in \(n\) and offers limited practical benefit, the high-temperature regime (\qtyrange{250}{300}{\kelvin}) provides a viable tuning mechanism, with \(\Delta n\) reaching up to \qty{0.006}{} at long wavelengths and \qty{0.02}{} near the band edge.


\bibliography{bib}

\end{document}